\documentclass[showpacs,prb,twocolumn]{revtex4}
\usepackage{amssymb}
\usepackage{epsfig}

\usepackage{amsmath}
\usepackage{graphicx}

\input{tcilatex}

\begin{document}

\title{Theory of enhancement of thermoelectric properties of materials with
nanoinclusions}
\author{Sergey V. Faleev}
\author{Fran\c{c}ois L\'{e}onard}
\email{fleonar@sandia.gov}
\affiliation{Sandia National Laboratories, Livermore, CA 94551}
\date{\today }

\begin{abstract}
Based on the concept of band-bending at metal/semiconductor interfaces as an
energy filter for electrons, we present a theory for the enhancement of the
thermoelectric properties of semiconductor materials with metallic
nanoinclusions. We show that the Seebeck coefficient can be significantly
increased due to a strongly energy-dependent electronic scattering time. By
including phonon scattering, we find that the enhancement of $ZT$ due to
electron scattering is important for high doping, while at low doping it is
primarily due to decrease of the phonon thermal conductivity.
\end{abstract}

\maketitle


\section{Introduction}

\label{sec1}

The energy conversion efficiency of thermoelectric devices depends on the
figure of merit $ZT=S^{2}\sigma T/\kappa $, where $S$, $\sigma $, $T$, and $%
\kappa $ are the Seebeck coefficient, electrical conductivity, temperature,
and thermal conductivity. In the best thermoelectric materials $ZT$ is
typically $\sim 1$, and it is difficult to increase $ZT$ beyond this value
because of competing effects of electrical and thermal conductivities.
Advances over the past decade show that it is possible to enhance $ZT$ in
nanostructured thin-film systems by taking advantage of quantum confinement
to enhance the power factor $S^{2}\sigma ,$\cite{Harman02} or to increase
phonon scattering at interfaces to reduce the lattice contribution to $%
\kappa $.\cite{Ven01} On the other hand, many existing and envisioned
thermoelectric applications will require a material that is itself of
macroscopic dimension. Therefore, recent reports of property enhancement in
bulk alloys possessing nanometer-scale compositional modulations have
generated much excitement. \cite{Hsu04,Lin05,Heremans05,Kim06} $ZT$ values
as high as 2.2 have been reported \cite{Hsu04,Lin05} in the (PbTe)$_{x}$%
(AgSbTe$_{2}$)$_{1-x}$ system, and have been ascribed to a large Seebeck
coefficient and low lattice thermal conductivity due to nanoscale clustering
of Ag and Sb. Heremans et al. \cite{Heremans05} reported that the Seebeck
coefficient in bulk PbTe can be increased significantly by precipitating a
fine distribution of Pb nanoinclusions, and suggested heuristically that the
increase in Seebeck coefficient originates from an energy-filtering effect
due to a strongly energy-dependent electronic scattering time. Kim et al.%
\cite{Kim06} observed an enhancement of the thermoelectric properties when
ErAs nanoparticles of 2.4 nm average diameter were embedded in a InGaAs
matrix, and ascribed the increase to a reduction in the phonon thermal
conductivity. Given these observations, a general understanding of the role
of nanoinclusions in enhancing the thermoelectric properties of materials is
needed, in particular to assess the relative importance of electronic and
phonon scattering.

In this paper we present a theoretical model and numerical calculations of
the thermoelectric properties of bulk semiconductors containing metallic
nanoparticles. Our model considers scattering of electrons on the
band-bending at the interfaces between the semiconductor host and randomly
distributed metallic islands. This causes energy-dependent scattering of
electrons, leading to an energy filterting effect that increases the Seebeck
coefficient. This provides an explicit physical model for the proposed
energy filtering effect.\cite{Heremans05} By combining this model with a
model for phonon scattering on the nanoinclusions, we predict significant
enhancement of the ZT factor.

We point out that while the role of metallic nanoinclusions may appear at
first to be similar to that of point defects for which extensive work has
been done, the physics is actually quite different. Indeed, in addition to
the electronic scattering, the phonon scattering on nanoinclusions occurs in
a completely different regime than that on point defects, as will be
discussed in section IVB.

The central idea in this paper is illustrated in Figure 1. There, spherical
metallic nanoinclusions with radius $R$ and volume fraction $x$ are randomly
distributed in a bulk semiconductor material. In general, at such
metal/semiconductor interfaces, charge transfer between the metal and the
semiconductor leads to band-bending away from the interface, characterized
by the electrostatic potential $V(r)$ (Fig. 1b). The presence of this
potential causes \textit{energy-dependent} scattering of electrons, as
illustrated in Fig. 1c. The high-energy electrons are unaffected by the
potential, but the low energy electrons can be strongly scattered. Because
the Seebeck coefficient depends on the energy derivative of the relaxation
time $d\ln \tau \left( E\right) /dE$ at the Fermi energy, this type of
energy filtering is precisely the prescription to increase the Seebeck
coefficient of thermoelectric materials.

\begin{figure}[h]
\centering
\epsfig{file=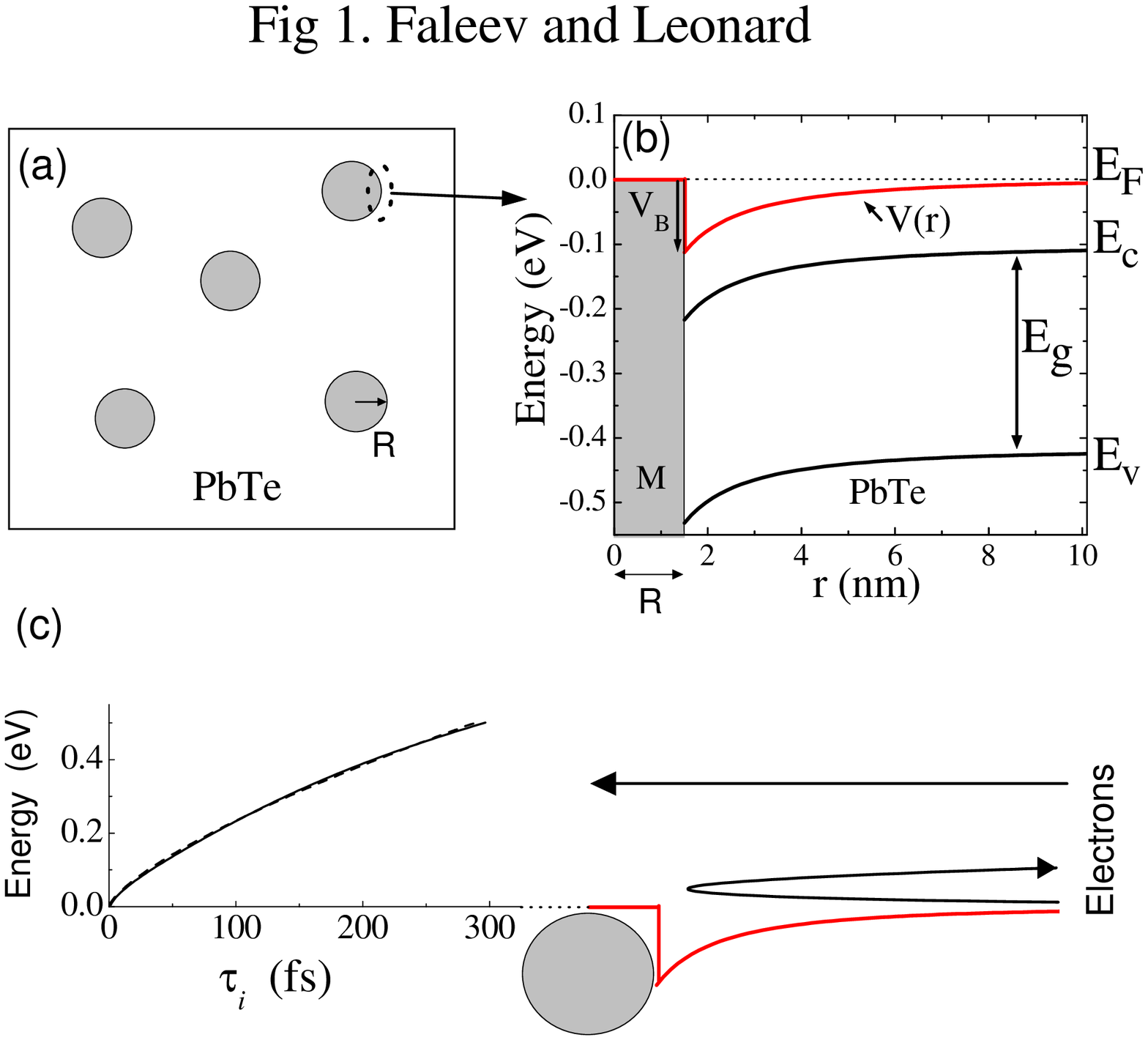,clip=,width=240pt,height=210pt}
\caption{(a) Schematic of the
semiconductor host with metallic nanoinclusions. Panel (b) shows an example
of the calculated potential $V(r)$ and the energy diagram for PbTe at $T=300K
$, $n=2.5\times 10^{19}cm^{-3}$, $V_{B}=-0.11eV$, and $R=1.5nm$. Panel (c)
illustrates the concept of energy filtering: low energy electrons scatter
strongly with the potential, but high energy electrons are unaffected. The
calculated electronic relaxation time for the potential of panel (b) is also shown.}
\end{figure}

Our theoretical model is based on the Boltzmann Transport Equation (BTE)
within the relaxation time approximation. We apply the model to a system of $%
n$-doped PbTe with metallic nanoinclusions because of the availability of
experimental data and good understanding of scattering mechanisms in bulk
PbTe \cite{RavichT71,RavichBook,Zayachuk97}, although the theory can be used
for any thermoelectric material.

\section{Charge and heat transport in bulk PbTe}

\label{sec2}

In this section we will review the expressions\cite%
{RavichT71,RavichBook,Zayachuk97} for the charge and heat transport in bulk
PbTe with $n$-type doping. The valence band of PbTe contains four energy
minima located at the $L$ points. The energy dispersion relation near each
minima is usually described by the Kane model\cite{RavichBook}%
\begin{equation}
\frac{\hbar ^{2}k_{l}^{2}}{2m_{l}^{\ast }}+\frac{\hbar ^{2}k_{t}^{2}}{%
m_{t}^{\ast }}=E(1+E/E_{g}),  \label{one}
\end{equation}%
where $E_{g}$ is the direct energy gap of PbTe, $\hbar $ is the Planck
constant, and $k$ and $m^{\ast }$ are the electron wavevector and effective
mass (at minimum energy point $k=0$, $E=0$) along the longitudinal (suffix $l
$) and transverse (suffix $t$) directions of the corresponding $L$ point.
For $n$-type PbTe the electron concentration is given by%
\begin{equation}
n[E_{F}]=\frac{(2m_{d}^{\ast }k_{B}T)^{3/2}}{3\pi ^{2}\hbar ^{3}}%
\int\limits_{0}^{\infty }\gamma (z)^{3/2}\left( -\frac{\partial f_{0}}{%
\partial z}\right) dz,  \label{two}
\end{equation}%
where $m_{d}^{\ast }=4^{2/3}(m_{l}^{\ast }m_{t}^{\ast 2})^{1/3}$ is the
density of states effective mass in which the 4-fold degeneracy is included, 
$k_{B}$ is the Boltzmann constant, $f_{0}(z)$ is the Fermi function written
in terms of dimensionless variables $z=E/(k_{B}T)$ and $z_{F}=E_{F}/(k_{B}T)$%
. $E_{F}$ is the Fermi energy, and the function $\gamma (z)=z+bz^{2}$, where 
$b=k_{B}T/E_{g}$. In the relaxation time approximation the BTE expressions
for electrical conductivity, $\sigma $, Seebeck coefficient, $S$, and
electron contribution to thermal conductivity, $\kappa _{e}$, are\cite%
{RavichBook} 
\begin{equation}
\sigma =\frac{e^{2}}{m_{c}^{\ast }}\frac{(2m_{d}^{\ast }k_{B}T)^{3/2}}{3\pi
^{2}\hbar ^{3}}\left\langle \tau (z)\right\rangle ,  \label{gh}
\end{equation}%
\begin{equation}
S=\frac{k_{B}}{e}\frac{\left\langle \tau (z)(z-z_{F})\right\rangle }{%
\left\langle \tau (z)\right\rangle },\hspace{0.03in}  \label{s}
\end{equation}%
and%
\begin{equation}
\kappa _{e}=\sigma T\frac{k_{B}^{2}}{e^{2}}\left( \frac{\left\langle \tau
(z)z^{2}\right\rangle }{\left\langle \tau (z)\right\rangle }-\left[ \frac{%
\left\langle \tau (z)z\right\rangle }{\left\langle \tau (z)\right\rangle }%
\right] ^{2}\right) ,  \label{kap_e}
\end{equation}%
where $m_{c}^{\ast }=3/(1/m_{l}^{\ast }+2/m_{t}^{\ast })$ is the effective
conductivity mass, and the average is defined as 
\begin{equation}
\left\langle A(z)\right\rangle \equiv \int\nolimits_{0}^{\infty }\frac{%
\gamma (z)^{3/2}}{1+2bz}\left( -\frac{\partial f_{0}}{\partial z}\right)
A(z)dz.  \label{aver}
\end{equation}

In bulk PbTe at room temperature the dominant contributions to the total
relaxation time, $\tau _{bulk}(z)$, are scattering by the deformation
potential of acoustic and optical phonons, and polar scattering by optical
phonons.\cite{RavichT71,RavichBook,Zayachuk97} We also take into account
scattering on the short-range potential of vacancies although it gives a
much smaller contribution compared to scattering by phonons. Thus, the total
relaxation time for bulk PbTe is given by 
\begin{equation}
\frac{1}{\tau _{bulk}(z)}=\frac{1}{\tau _{PO}(z)}+\frac{1}{\tau _{a}(z)}+%
\frac{1}{\tau _{o}(z)}+\frac{1}{\tau _{v}(z)}.  \label{tm1}
\end{equation}

The relaxation time due to polar scattering by optical phonons reads\cite%
{RavichBook,Zayachuk97}%
\begin{equation}
\tau _{PO}(z)=\frac{\hbar ^{2}\gamma (z)^{1/2}F^{-1}}{e^{2}(2m_{d1}^{\ast
}k_{B}T)^{1/2}(\epsilon _{\infty }^{-1}-\epsilon _{0}^{-1})\gamma ^{\prime
}(z)},  \label{five}
\end{equation}%
where $m_{d1}^{\ast }=(m_{l}^{\ast }m_{t}^{\ast 2})^{1/3}$ is the density of
states effective mass in the single valley, the function $\gamma ^{\prime
}(z)=1+2bz$, $\epsilon _{0}$ and $\epsilon _{\infty }$ are the static and
high frequency permittivities, and 
\begin{eqnarray}
F &=&1-\delta \ln (1+\frac{1}{\delta })-\frac{2bz(1+bz)}{(1+2bz)^{2}}
\label{eii} \\
&&\times \lbrack 1-2\delta +2\delta ^{2}\ln (1+\frac{1}{\delta })].  \notag
\label{seven}
\end{eqnarray}%
Here $\delta =(2kr_{0})^{-2}$ with $r_{0}$ the screening length of optical
phonons:%
\begin{eqnarray}
r_{0}^{-2} &=&\frac{2^{5/2}e^{2}m_{d}^{\ast 3/2}(k_{B}T)^{1/2}}{3\pi
^{2}\hbar ^{3}}  \label{fivep} \\
&&\times \int\limits_{0}^{\infty }\gamma (z)^{1/2}\gamma ^{\prime }(z)\left(
-\frac{\partial f_{0}}{\partial z}\right) dz.  \notag
\end{eqnarray}

The relaxation time due to scattering by the deformation potential of
acoustic and optical phonons, and also due to scattering on the short range
potential of vacancies can be written generally as\cite%
{RavichBook,Zayachuk97}%
\begin{equation}
\tau _{m}(z)=\frac{\tau _{0,m}}{\gamma (z)^{1/2}\gamma ^{\prime
}(z)[(1-A)^{2}-B]},  \label{nine}
\end{equation}%
where $A=bz(1-K_{m})/\gamma ^{\prime }(z)$, and $B=8bz(1+bz)K_{m}/(3\gamma
^{\prime 2}(z)),$ with the suffix $m=a$ for acoustic phonons, $m=o$ for
optical phonons, and $m=v$ for vacancies. The constants $\tau _{0,m}$ and $%
K_{m}$ are defined as 
\begin{eqnarray}
\tau _{0,a} &=&\frac{2\pi \hbar ^{4}C_{l}}{E_{ac}^{2}(2m_{d1}^{\ast
}k_{B}T)^{3/2}},\quad K_{a}=\frac{E_{av}}{E_{ac}},  \label{ten} \\
\tau _{0,o} &=&\frac{2\hbar ^{2}a^{2}(\hbar \omega _{0})^{2}\rho }{\pi
E_{oc}^{2}(2m_{d1}^{\ast }k_{B}T)^{3/2}},\quad K_{o}=\frac{E_{ov}}{E_{oc}},
\label{ov}
\end{eqnarray}%
and%
\begin{equation}
\tau _{0,v}=\frac{\pi \hbar ^{4}}{U_{vc}^{2}m_{d1}^{\ast }(2m_{d1}^{\ast
}k_{B}T)^{1/2}N_{v}},\quad K_{v}=\frac{U_{vv}}{U_{vc}}.  \label{vv}
\end{equation}%
Here $C_{l}$ is a combination of elastic constants, $E_{ac}$ and $E_{av}$
are the acoustic phonon deformation potential coupling constants for
conduction and valence bands, $E_{oc}$ and $E_{ov}$ are optical phonon
deformation potential coupling constants for conduction and valence bands, $%
U_{vc}$ and $U_{vv}$ are coupling constants of the short range potential of
vacancies for conduction and valence bands, $a$ is the lattice constant, $%
\omega _{0}$ is the frequency of optical phonons, and $\rho $ is the mass
density. $N_{v}$ is the concentration of vacancies calculated from the
condition that one vacancy gives two charge carriers, $N_{v}=n/2$.

\ The parameters used for calculation of the relaxation times in bulk PbTe
at $T=300K$ are taken from Ref. [\onlinecite{Zayachuk97}]. These parameters
are shown in Table \ref{T1}. For calculations at different temperatures we
assumed the values of these parameters to be the same as for $T=300K$ except
for $E_{g}$ and $m_{t}^{\ast }$ which were linearly interpolated and
extrapolated using $T=4.6K$ and $T=300K$ values,\cite{Zayachuk97} with $%
E_{g} $ saturating for $T>400K$.\cite{RavichBook}

\begin{table}[h] \centering%
\begin{tabular}{|l|c|l|c|}
\hline\hline
Parameter & Value & Parameter & Value \\ \hline\hline
$E_{g}$ & 0.315 eV & $\hbar \omega _{0}$ & 0.0136 eV \\ \hline
$m_{t}^{\ast }/m_{0}$ & 0.0453 & $a$ & 6.461 \AA \\ \hline
$m_{l}^{\ast }/m_{0}$ & 0.24 & $\rho $ & 8.24 g/cm \\ \hline
$\epsilon _{0}$ & 400 & $E_{ac}$ & 15 eV \\ \hline
$\epsilon _{\infty }$ & 32.6 & $E_{oc}$ & 26 eV \\ \hline
$C_{l}$ & 7.1$\times 10^{10}$N/m & $K_{a,o}$ & 1.5 \\ \hline
$U_{vc}$ & 3$\times 10^{-34}$erg cm$^{3}$ & $K_{v}$ & 1.5 \\ \hline
\end{tabular}%
\caption{Parameters used to calculate the relaxation times for bulk PbTe at
T=300K.\cite{Zayachuk97} $m_0$ is the free electron mass.}\label{T1}%
\end{table}%

\section{Electron scattering on band-bending potential of nanoinclusions}

\subsection{Band-bending potential}

In our model we assume that spherical metallic nanoinclusions with radius $R$
and volume fraction $x$ are randomly distributed in a $n$-doped PbTe host
material. In this section we will calculate the contribution to the
relaxation time due to scattering of electrons on the band-bending potential
at the metal-semiconductor interface.

For a \textit{single} nanoinclusion, the electrostatic potential $V(r)$ can
be calculated by solving the Poisson equation 
\begin{equation}
\frac{\epsilon _{0}}{4\pi e^{2}}\frac{1}{r}\frac{d^{2}}{dr^{2}}%
rV(r)=n[E_{F}]-n[E_{F}-V(r)].  \label{spV}
\end{equation}%
The right-hand-side of this expression is simply the spatially varying
charge [see Eq. (\ref{two})] calculated by assuming a rigid shift of the
electronic bands with the local potential $V(r)$. We solve the Poisson
equation with the boundary conditions $V\left( \infty \right) =0$ and $%
V(R)=V_{B}$. ($V_{B}$ is the potential at the semiconductor/metal interface.
The value of $V_{B}$ is fixed for a particular metal, and depends on the
detailed properties of the interface. However, one may consider it to be an
optimization parameter provided that the physics of the metal/semiconductor
interface allows tailoring of $V_{B}$ by choosing the metal.) We used the
fourth-order Runge-Kutta and shooting methods in order to solve Eq. (\ref%
{spV}). Figure 1b shows an example of the calculated potential $V(r)$.

For small values of $V_{B}$ or for large $r$ (when the potential is screened
and small) the right-hand-side of Eq. (\ref{spV}) can be linearized with
respect to small $V$: 
\begin{equation}
\frac{d^{2}}{dr^{2}}rV(r)=\frac{1}{\lambda ^{2}}rV(r),  \label{ex}
\end{equation}%
where $\lambda $ is the screening length. The solution of Eq. (\ref{ex}) is 
\begin{equation}
V(r)\propto \frac{1}{r}e^{-r/\lambda }.  \label{VV}
\end{equation}%
For degenerate electrons the expression for $\lambda $ takes the simple form 
\begin{equation}
\frac{1}{\lambda ^{2}}=\frac{2e^{2}(2m_{d}^{\ast })^{3/2}}{\epsilon _{0}\pi
\hbar ^{3}}\left( E_{F}+\frac{E_{F}^{2}}{E_{g}}\right) ^{1/2}\left( 1+2\frac{%
E_{F}}{E_{g}}\right) .  \label{lam0}
\end{equation}%
Due to the large value of the dielectric constant in PbTe, $\epsilon _{0}=400
$, the screening length for typical doping concentrations is several times
larger then the wavelength of electrons on the Fermi surface. For example,
for $n=5\times 10^{19}$cm$^{-3}$ one can obtain from Eq. (\ref{lam0}) $%
\lambda =11$nm and $k_{F}\lambda =7$, where $k_{F}$ is the wavevector of the
electron on the Fermi surface. This value of $k_{F}\lambda $ slowly varies
with doping (for degenerate electrons we can use Eqs (\ref{two}) and (\ref%
{lam0}) to obtain the dependence on doping $k_{F}\lambda \propto n^{1/6}$).
On the other hand, a large value of $\lambda $ can lead to an overlap of the
band-bending between nanoparticles, which may change the bulk carrier
concentration. We have restricted our calculations to a parameter range
(doping and inclusion volume fraction) where such effects are not
significant.

\subsection{Relaxation time for scattering on nanoinclusions}

When the electron scattering on nanoinclusions is taken into account the
total\textit{\ }relaxation time $\tau $ is 
\begin{equation}
\tau ^{-1}=\tau _{bulk}^{-1}+\tau _{i}^{-1},  \label{ttot}
\end{equation}%
where the relaxation time for bulk PbTe is given by Eq. (\ref{tm1}), and $%
\tau _{i}$ is the relaxation time due to scattering by $V(r)$ at randomly
distributed metallic inclusions 
\begin{equation}
\tau _{i}^{-1}=n_{i}v\sigma _{t}.  \label{ti}
\end{equation}%
Here 
\begin{equation}
n_{i}=3x/(4\pi R^{3})  \label{ni}
\end{equation}%
is the concentration of inclusions, $\sigma _{t}$ is the electronic
transport scattering cross-section, and $v=\partial _{p}E_{p}$ is the
electronic velocity with $p$ the momentum.

In order to calculate the transport cross-section in a system with
nonparabolic energy dispersion, we consider an electron with momentum $%
\mathbf{p}$ and\ wave function $\psi _{\mathbf{p}}(\mathbf{r})=u_{\mathbf{p}%
}(\mathbf{r})e^{i\mathbf{pr/}\hbar }$ in the periodic field of the
unperturbed PbTe crystal of unit volume. Here $u_{\mathbf{p}}(\mathbf{r})$
is the periodic Bloch amplitude. As mentioned earlier, the nonparabolicity
of the electron energy dispersion near the conduction band minima is usually
described by the Kane model\cite{RavichBook} 
\begin{equation}
E_{p}(1+E_{p}/E_{g})\hspace{0.03in}=p^{2}/2m_{d1}^{\ast }.  \label{dis1}
\end{equation}%
The isotropic energy dispersion in a form of Eq. (\ref{dis1}) with density
of state mass $m_{d1}^{\ast }$ is usually assumed in the calculation of the
relaxation time.\cite{RavichBook,RavichT71,Zayachuk97} The transition
probability for scattering from state $\psi _{\mathbf{p}_{i}}$ to state $%
\psi _{\mathbf{p}_{f}}$ per unit time due to a perturbation potential $V(r)$
is given by the standard formula of perturbation theory\cite{Landau76}%
\begin{eqnarray}
dw_{fi} &=&\frac{2\pi }{\hbar }\left| V_{\mathbf{p}_{f}\mathbf{p}_{i}}+\int 
\frac{V_{\mathbf{p}_{f}\mathbf{p}_{1}}V_{\mathbf{p}_{1}\mathbf{p}_{i}}}{%
E_{p_{i}}-E_{p_{1}}}d\nu _{1}+...\right| ^{2}  \notag \\
&&\times \delta (E_{p_{f}}-E_{p_{i}})d\nu _{f},  \label{FGR2}
\end{eqnarray}%
where $d\nu =d^{3}\mathbf{p}/(2\pi \hbar )^{3}$. The matrix elements are%
\begin{equation}
V_{\mathbf{p}^{\prime }\mathbf{p}}\equiv \int \psi _{\mathbf{p}^{\prime
}}^{\ast }(\mathbf{r})V(\mathbf{r})\psi _{\mathbf{p}}(\mathbf{r})d^{3}%
\mathbf{r\approx }\int e^{i(\mathbf{p}-\mathbf{p}^{\prime })\mathbf{r/}\hbar
}V(\mathbf{r})d^{3}\mathbf{r},  \label{vpp}
\end{equation}%
where we used the fact that collisions with a small momentum transfer
dominate scattering on the slow varying bend-bending potential, therefore
the Bloch amplitudes entering Eq. (\ref{vpp}) are rather close to each other
and the overlap factor is about unity $\int u_{\mathbf{p}^{\prime }}^{\ast }(%
\mathbf{r})u_{\mathbf{p}}(\mathbf{r})d^{3}\mathbf{r\approx }1$.

Applying Eq. (\ref{vpp}) to Eq. (\ref{FGR2}), the calculation of $dw_{fi}$
becomes identical to the calculation of the transition probability for
scattering of a plane wave $e^{i\mathbf{p}_{i}\mathbf{r/}\hbar }$ in a model
system described by an equation 
\begin{equation}
(E_{\hat{p}}+V)\psi =E_{p}\psi   \label{seq1}
\end{equation}%
with unperturbed Hamiltonian $E_{\hat{p}}$, $\mathbf{\hat{p}}\equiv -i\hbar
\partial _{\mathbf{r}}$, and perturbation potential $V(\mathbf{r}).$
Applying the operator $1+E_{\hat{p}}/E_{g}$ to Eq. (\ref{seq1}) one obtains 
\begin{equation}
\frac{\hat{p}^{2}}{2m_{d1}^{\ast }}\psi =\left[ \frac{p^{2}}{2m_{d1}^{\ast }}%
+\frac{V^{2}}{E_{g}}-V\left( 1+2\frac{E_{p}}{E_{g}}\right) \right] \psi .
\label{seq2}
\end{equation}%
Here we neglect the commutator term $E_{g}^{-1}[E_{\hat{p}},V(r)]\psi $ 
\begin{equation}
E_{g}^{-1}[E_{\hat{p}},V(r)]\mathbf{\approx }\frac{E_{F}}{E_{g}}\frac{1}{%
(k_{F}\lambda )^{2}}V(r)\ll V(r)  \label{eq100}
\end{equation}%
using the fact that $rV(r)$ is a slow varying function and $k_{F}\lambda \gg
1$.

Eq. (\ref{seq2}) has the form of the usual Schr\"{o}dinger equation that can
be used for numerical solution of the scattering problem with a potential 
\begin{equation}
U_{p}(r)=V(r)\left( 1+2\frac{E_{p}}{E_{g}}\right) -\frac{V^{2}(r)}{E_{g}}.
\label{ppot}
\end{equation}

\subsection{Calculation of transport cross-section}

The transport cross-section for scattering on the spherically symmetric
potential is given by\cite{Landau76}%
\begin{equation}
\sigma _{t}=2\pi \int\limits_{0}^{\pi }|f(\theta )|^{2}(1-\cos \theta )\sin
\theta d\theta ,  \label{Eq14}
\end{equation}%
where $f(\theta )$ is the scattering amplitude defined from the large-$r$
asymptotic of the wave function%
\begin{equation}
\psi \approx e^{i\mathbf{pr/}\hbar }+\frac{f(\theta )}{r}e^{ipr/\hbar }.
\label{Eq15}
\end{equation}%
The wave function $\psi $ is a solution of the Schr\"{o}dinger equation (\ref%
{seq2}) with potential $U_{p}(r)$ given by Eq. (\ref{ppot}). It can be
expressed as a sum of contributions with different angular momentum $l$\cite%
{Landau76} 
\begin{equation}
\psi =\sum\limits_{l=0}^{\infty }P_{l}(\cos \theta )R_{kl}(r),  \label{Eq16}
\end{equation}%
where $P_{l}$ are the Legendre polynomials and $R_{kl}(r)$ are solutions of
the radial Schr\"{o}dinger equation $(k=p/\hbar )$:

\begin{equation}
\frac{1}{r^{2}}\partial _{r}r^{2}\partial _{r}R_{kl}+\left( k^{2}-\frac{%
l(l+1)}{r^{2}}-\frac{2m_{d1}^{\ast }}{\hbar ^{2}}U_{p}(r)\right) R_{kl}=0.
\label{Eq17}
\end{equation}%
The large-$r$ asymptotic of $R_{kl}(r)$ has the form 
\begin{equation}
R_{kl}(r)\propto \sin (kr-l\pi /2+\delta _{l})/r,  \label{Eq18}
\end{equation}%
where $\delta _{l}$ is the phase shift of $R_{kl}(r)$ relative to the
potential-free solution. The scattering amplitude $f(\theta )$ can be
expressed in terms of the phase shifts as\cite{Landau76}

\begin{equation}
f(\theta )=\frac{1}{2ik}\sum_{l=0}^{\infty }(2l+1)P_{l}(\cos \theta
)(e^{2i\delta _{l}}-1).  \label{fl}
\end{equation}

The calculation of the transport cross-section (\ref{Eq14}) with scattering
amplitude (\ref{fl})\ can be performed in the same way\cite{Landau76} as the
usual cross-section [without the $(1-\cos \theta )$ factor in Eq. (\ref{Eq14}%
)]. In the integrals $\int\limits_{0}^{\pi }P_{l}(\cos \theta )P_{l^{\prime
}}(\cos \theta )(1-\cos \theta )\sin \theta d\theta \hspace{0.03in}$that
appear in the evaluation of the transport cross-section (\ref{Eq14}) only
terms with $l^{\prime }=l,l\pm 1$ give nonvanishing contributions.\cite%
{Landau76} After integration over $\theta ,$ Eq. (\ref{Eq14}) can be
expressed in terms of the $\delta _{l}$ as%
\begin{equation}
\sigma _{t}=\frac{4\pi }{k^{2}}\sum\limits_{l=1}^{\infty }l\sin ^{2}(\delta
_{l}-\delta _{l-1}).  \label{Eq19}
\end{equation}

We used the following numerical procedure to solve the Schr\"{o}dinger
equation (\ref{Eq17}) and calculate the phase shifts. The spherical Bessel
function $j_{l}(kr)$ is a regular solution of Eq. (\ref{Eq17}) for $r<R$, in
the region where $U_{p}(r)=0$. We used the fourth-order Runge-Kutta method
to solve Eq. (\ref{Eq17}) for $r>R$ with boundary conditions such that the
solution $R_{l}(kr)$ and its derivative match the spherical Bessel function
at $r=R$ . At some large $r=r_{\max }$ we assume that the potential $U_{p}$
vanishes and match the solution $R_{l}(kr)$ and its derivative $%
R_{l}^{\prime }(kr)$ to a linear combination of the spherical Bessel
function $j_{l}(kr)$ and spherical Neumann function $y_{l}(kr)$ (which is
another solution of Eq. (\ref{Eq17}) for $U_{p}(r)=0$):%
\begin{equation}
R_{l}(kr)|_{r=r_{\max }}\rightarrow \alpha j_{l}(kr)+\beta y_{l}(kr),
\end{equation}%
with $\alpha =(R_{l}y_{l}^{\prime }-R_{l}^{\prime
}y_{l})/(j_{l}y_{l}^{\prime }-j_{l}^{\prime }y_{l})$ and $\beta
=-(R_{l}j_{l}^{\prime }-R_{l}^{\prime }j_{l})/(j_{l}y_{l}^{\prime
}-j_{l}^{\prime }y_{l})$. Finally the phase shift $\delta _{l}$ is given by 
\begin{equation}
\delta _{l}=-\arctan (\beta /\alpha ).  \label{psh}
\end{equation}

In conjunction with equations (\ref{ti}) and (\ref{Eq19}) this provide
expressions to numerically calculate $\tau _{i}$ for a given $V(r)$. An
example of the numerically calculated $\tau _{i}$ is given in Fig. 1c for
scattering by $V(r)$ of Fig. 1b with inclusion volume fraction $x=5\%$. A
simple fit gives the dependence $\tau _{i}\left( E\right) \sim E^{1.39}$ as
shown by the dashed line in Fig. 1c. This energy dependence of $\tau _{i}$
is much stronger than that of $\tau _{bulk}$, and leads to enhancement of
the Seebeck coefficient.

To obtain an analytical description of $\tau _{i}\left( E\right) $, we also
calculated $\tau _{i}$ in the Born approximation by using Fermi's golden
rule [first term in r.h.s. of Eq. (\ref{FGR2})] 
\begin{equation}
\frac{1}{\tau _{i}^{Born}}=\frac{p^{2}}{2\pi \hbar ^{4}}\frac{dp}{dE}%
\int\limits_{0}^{\pi }|V_{\mathbf{p}^{\prime }\mathbf{p}}|^{2}(1-\cos \theta
)\sin \theta d\theta .  \label{Eq108}
\end{equation}%
Here $\theta $ is the angle between initial and final momenta $\mathbf{p}$
and $\mathbf{p}^{\prime }$. The expression for $\tau _{i}^{Born}$ can be
simplified by taking the angle integration in Eq. (\ref{Eq108}) for $V_{%
\mathbf{p}^{\prime }\mathbf{p}}$ and making the substitutions of integration
variables $t=2kR\sin \frac{\theta }{2}$ and $y=r/R$. Finally one obtains 
\begin{equation}
\tau _{i}^{Born}\left( E\right) =E^{3/2}\frac{(1+E/E_{g})^{3/2}}{1+2E/E_{g}}%
\frac{R}{x}\frac{4\sqrt{2m_{d1}^{\ast }}}{3\alpha (E,R)},  \label{Eq109}
\end{equation}%
where 
\begin{equation}
\alpha (E,R)=\int_{0}^{2kR}\left| \tint\nolimits_{1}^{\infty }\sin
(yt)V(yR)ydy\right| ^{2}tdt.  \label{Eq110}
\end{equation}%
Numerical tests show that for $|V_{B}|\lesssim 0.1eV$ the Born approximation
is valid, $\tau _{i}(E)\approx \tau _{i}^{Born}(E)$, while for $%
|V_{B}|>0.1eV $, $\tau _{i}^{Born}\left( E\right) $ begins to deviate from $%
\tau _{i}(E)$ calculated from the exact solution of Schr\"{o}dinger's
equation (\ref{Eq17}). Nevertheless, Eq. (\ref{Eq109})\ allows us to analyze
the energy dependence of the relaxation time that is difficult to do by
using the exact formulas (\ref{ti}) and (\ref{Eq19}). For energies $E\gtrsim
0.1eV$, the integral over variable $t$ in (\ref{Eq110}) weakly depends on
the upper limit of the integration, and the function $\alpha (E,R)$ varies
slowly with both $E$ and $R$. Thus, we have $\tau _{i}^{Born}\left( E\right)
\sim E^{3/2} $, in good agreement with the full numerical calculations which
yielded a dependence $E^{1.39}$ . Comparing the result $\tau
_{i}^{Born}\left( E\right) \sim E^{3/2}$ with the expression (\ref{ti}) (and
using $v(E)\sim \sqrt{E}$) we find that the electronic scattering
cross-section of the band-bending potential depends on energy as $E^{-2}$;
this strong energy dependence is responsible for the superlinear energy
dependence of $\tau _{i}\left( E\right) $.

\section{Results and discussion}

\subsection{Enhancement of the Seebeck coefficient and power factor}

The calculation of the total relaxation time allows us to obtain $S$, $%
\sigma $ and $\kappa _{e}$ using the expressions (\ref{gh}-\ref{kap_e}). We
first consider a specific case by adopting a simple model for the interface
potential $V_{B}=\Phi _{m}-\chi +E_{F}$ with $\Phi _{m}$ the metal
workfunction and $\chi $ the electron affinity, and choose $\Phi _{m}-\chi =$
$-0.35eV$ corresponding to Pb nanoinclusions (work function $\Phi
_{m}=4.25eV $ \cite{wpb}) and an electron affinity for PbTe $\chi
_{PbTe}=4.6eV$ \cite{xipbte}. Figure 2 shows the calculated room temperature
Seebeck coefficient as a function of the doping $n$ and fixed nanoinclusion
volume fraction $x=5\%$. We note the excellent agreement between the
experimentally measured $S$ (filled circles) \cite{RavEx} and that
calculated numerically (solid line) for bulk PbTe. In addition, one can see
that for any nanoinclusion radius, the Seebeck coefficient is always
increased compared to that of the inclusion-free system. In fact, for the
smallest nanoinclusion radius considered here ($1.5$nm), the enhancement in $%
S$ is over 100\% at high doping.

\begin{figure}[h]
\centering
\epsfig{file=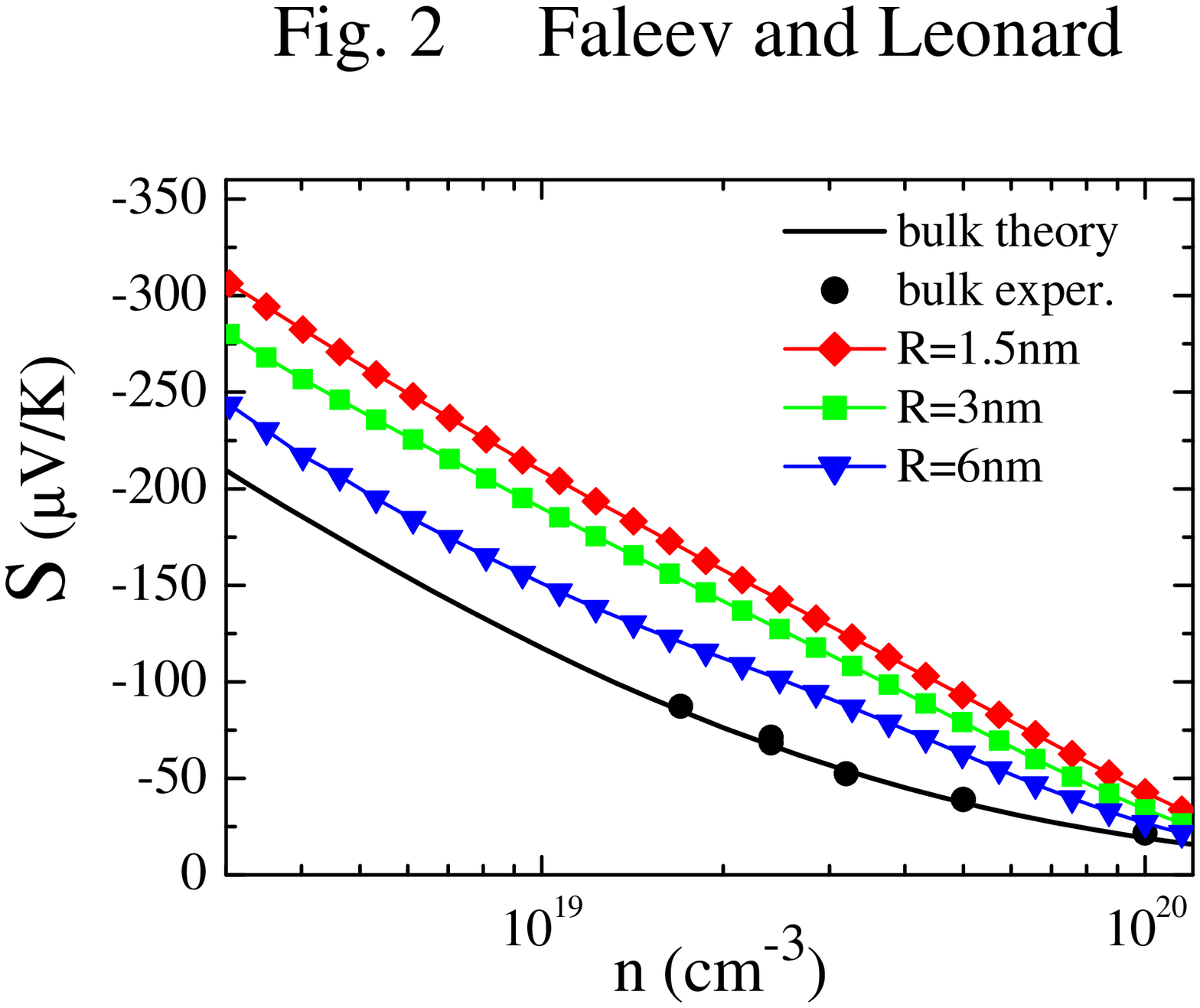,clip=,width=240pt,height=180pt}
\caption{Calculated Seebeck coefficient
for PbTe with metallic nanoinclusions as a function of the doping for
several different values of the nanoinclusion radius.}
\end{figure}

It is interesting to consider the impact of $V_{B}$ on the calculated
Seebeck coefficient. Figure 3a shows $S$ as a function of $V_{B}$. It is
clearly seen from this figure that the presence of an extended electrostatic
potential leads to an increase in $S$ regardless of the sign of $V_{B}$. ($%
V_{B}=0$ is equivalent to bulk PbTe with nanoinclusions. Negative values
correspond to the situation of Fig. 1, and positive values represent a
Schottky barrier). This general behavior can be understood (at least for
small $|V_{B}|$) from the Born approximation which predicts that the inverse
scattering time is proportional to the square of the perturbation potential.
With increase of $|V_{B}|$ the contribution to the total inverse relaxation
time from inclusion scattering increases, leading to an increase of $S$
because the energy dependence of $\tau $ changes from that of $\tau _{bulk}$
to the more strongly energy dependent $\tau _{i}$. For large values of $%
|V_{B}|$ the contribution of island scattering becomes dominant and $S$
saturates as seen in Fig. 3a. \ 

Figure 3a also shows the calculated values of $\sigma $ as a function of the
interface potential $V_{B}$. The conductivity decreases as $|V_{B}|$ is
increased, with a fairly symmetric behavior for $\pm V_{B}$. Combining the
results for $\sigma $ and $S$, we obtain the power factor $S^{2}\sigma $ as
depicted in Fig. 3b. There, one can see that the power factor is increased
compared to that at $V_{B}=0$, in a range of interface potentials $%
-0.15eV<V_{B}<0.15eV$. The power factor has two maxima at some optimal
values of $V_{B}$ because the Seebeck coefficient saturates for large $%
|V_{B}|$ while the electrical conductivity $\sigma $ continues to decrease
with increase of $|V_{B}|$.

\begin{figure}[h]
\centering
\epsfig{file=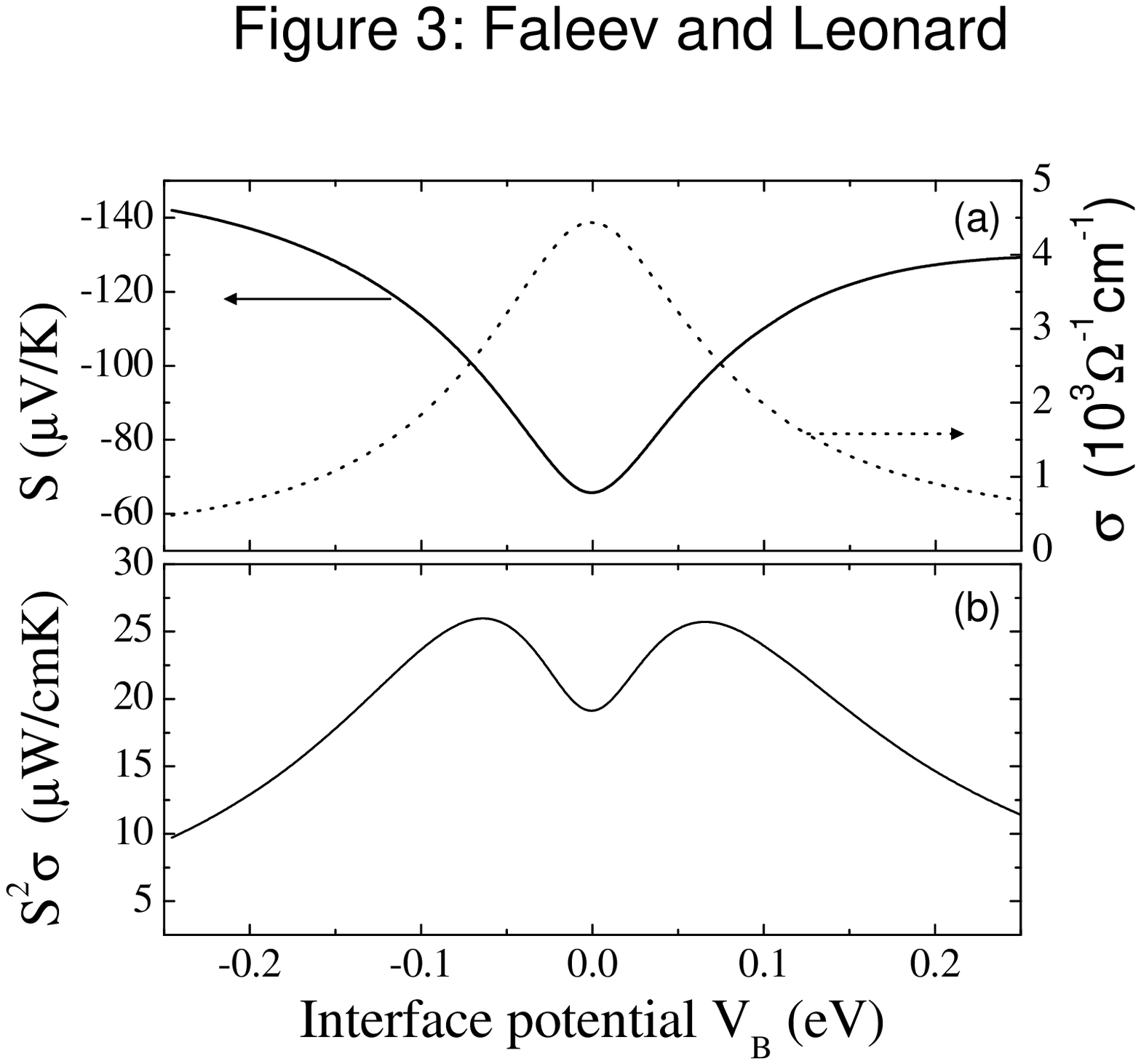,clip=,width=240pt,height=240pt}
\caption{Panel (a) shows the calculated Seebeck
coefficient and conductivity for PbTe as a function of the interface
potential $V_{B}$. Panel (b) shows the resulting power factor. Parameters
are $R=1.5nm$, $T=300K$, $x=5\%$, and $n=2.5\times 10^{19}cm^{-3}$.}
\end{figure} 

For the optimal interface potential $V_{B}\approx \pm 0.07eV$, we find that
the power factor is increased by $\sim 35\%$. Importantly, the power factor
does not decrease substantially over a wide range of values of the interface
potential. Thus, it is possible to take full advantage of the reduction in
thermal conductivity due to phonon scattering at the nanoinclusions, as we
will discuss later.

Instead of optimizing the value of the interface potential $V_{B}$ to
achieve the maximum power factor as shown in Figure 3b, we can keep $V_{B}$
fixed (by choosing a specific metal for the inclusions) and optimize other
parameters, for example the inclusion volume fraction or radius. To analyze
the dependence of the transport coefficients on these parameters we can use
Eq. (\ref{Eq109}). As we noted above, the function $\alpha (E,R)$ in Eq. (%
\ref{Eq110}) varies slowly with both $E$ and $R$, so the inverse relaxation
time due to electron scattering by inclusions can be approximated as $\tau
_{i}^{-1}(E)\approx CE^{-3/2}$, where the constant $C$ depends on $V_{B}$, $x
$, and $R$ mostly through the combination (at least for small $V_{B}$) 
\begin{equation}
C\propto V_{B}^{2}x/R.  \label{ccc}
\end{equation}%
In turn, the transport coefficients ($S$, $\sigma $, and $\kappa _{e}$)
depend only on this ratio of parameters. This means that if any two
parameters out of these three are fixed one can always adjust the third
parameter (for example, the one that can be most easily tuned in the
experiment) to maximize the power factor.

\begin{figure}[h]
\centering
\epsfig{file=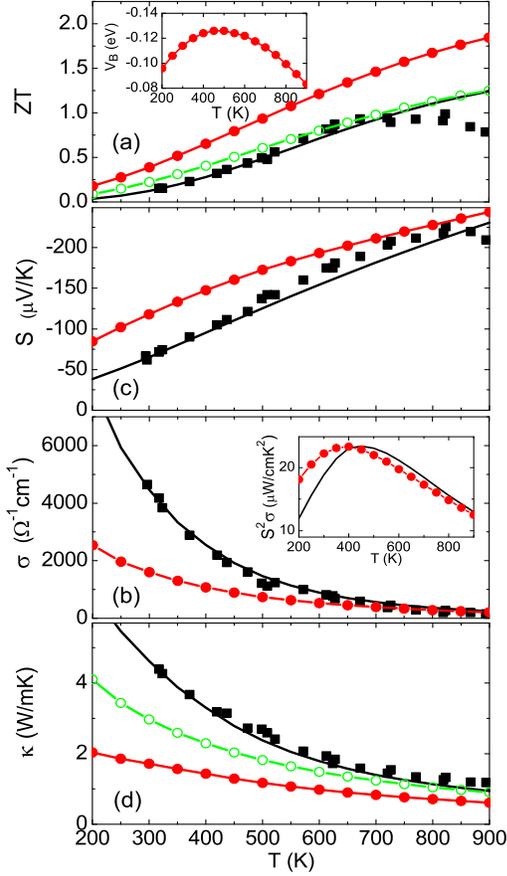,clip=,width=204pt,height=340pt}
\caption{(color online) Thermoelectric coefficients as a
function of temperature calculated for $n=2.5\times 10^{19}cm^{-3}$, $x=5\%$%
, and $R=1.5$ nm. (a) The optimized $ZT$ factor, with the optimal values of $%
V_{B}$ shown in the inset.(b) The Seebeck coefficient. (c) The electrical
conductivity. Inset shows optimized (filled circles) and bulk PbTe (solid
line) power factor $S^{2}\protect\sigma $. (d) The thermal conductivity. In
all panels, solid circles include electron and phonon scattering on the
inclusions; open circles include electron scattering on the inclusions but
with bulk PbTe values of $\protect\kappa _{ph}$; solid lines and filled
squares are the calculated and measured\protect\cite{Efimova71} values for
bulk PbTe.}
\end{figure}

Before closing this section, we remark that we have searched for resonant
tunneling states in the positive $V_{B}$ regime as a way to enhance $S$.
However, we have not found any significant increase in $S$ beyond that
already discussed. The reason is that, because the potential contains
contributions from several Legendre polynomials $l$, the appearance of a
resonant state for one value of $l$ is washed out by the nonresonant
conditions in the other channels.

\subsection{Enhancement of the ZT factor}

While $S$ and $\sigma $ are quantities of fundamental interest, for
applications it is usually $ZT=S^{2}\sigma T/\kappa $ that is most
important. The electronic relaxation time calculated above leads directly to
the electronic thermal conductivity $\kappa _{e}$. Since the total thermal
conductivity is the sum of electronic and phonon contributions, $\kappa
=\kappa _{e}+\kappa _{ph}$, to obtain $ZT$ we also need to calculate $\kappa
_{ph}$. For this purpose, we adopt a previous method \cite{Kim06,Kim06JAP}
that considered the scattering of phonons on nanoinclusions, with the
scattering mechanism for short wavelength phonons being the different sound
velocities in the host and nanoinclusions. This approach has been shown to
give excellent agreement with experiments on nanoscale ErAs inclusions in
InGaAs \cite{Kim06}. For $T\gtrsim T_{D}$ ($T_{D}=130K$ is the Debye
temperature of PbTe) $\kappa _{ph}$ can be written as \cite{RavichBook} 
\begin{equation}
\kappa _{ph}\approx \frac{k_{B}}{2\pi ^{2}v_{s}\hbar ^{3}}%
\tint\nolimits_{0}^{k_{B}T_{D}}\tau _{ph}\left( \hbar \omega \right)
^{2}d\left( \hbar \omega \right) ,  \label{eq18}
\end{equation}%
where $v_{s}$ is the speed of sound in PbTe and $\hbar \omega $ is the
phonon energy. The phonon relaxation time $\tau _{ph}$ is given by 
\begin{equation}
\tau _{ph}^{-1}=\tau _{U}^{-1}+\tau _{D}^{-1},  \label{pht}
\end{equation}%
where $\tau _{U}^{-1}=cT\omega ^{2}$ is the contribution of umklapp
scattering \cite{RavichBook} and $\tau _{D}$ is due to scattering by
nanoinclusions. The constant $c$ was determined from Eq. (\ref{eq18}) using
the experimental value $\kappa _{ph}^{bulk}=2.0W/mK$ for PbTe at $T=300K$.
For $\tau _{D}$ we used the expression derived in Refs. [
\onlinecite{Kim06,Kim06JAP}]. In the near geometrical scattering
regime ($qR\gtrsim 1$) $\tau _{D}$ reads 
\begin{equation}
\tau _{D}^{-1}=n_{i}v_{s}(2\pi R^{2})[1-\sin (2\xi )/\xi +\sin ^{2}(\xi
)/\xi ^{2}],  \label{eq19}
\end{equation}%
where $\xi =qR(v_{s}/v_{s}^{\prime }-1)$, $q$ is the phonon wave vector, and 
$v_{s}^{\prime }$ is the speed of sound inside the inclusion. Numerical
tests show that when the difference in the sound velocities is larger than
20\% the integrated quantity $\kappa _{ph}$ weakly depends on this
difference and $\tau _{D}$ can be approximated by its geometrical limit
value 
\begin{equation}
\tau _{D}^{-1}=n_{i}v_{s}(2\pi R^{2})=\frac{3}{2}\frac{x}{R}v_{s}.
\label{tphf}
\end{equation}%
Note that this phonon scattering regime is opposite to that on point defects
where $qR\ll 1$.

Figure 4 shows $ZT$ and its components calculated for $x=5\%$, $R=1.5nm$ and
a doping $n=2.5\times 10^{19}cm^{-3}$, as a function of temperature. We
discuss this doping first because experimental values of $ZT$ and all of its
components are readily available for inclusion-free PbTe, and can be
compared with our calculations; indeed, the calculated values for $ZT$, $S$, 
$\sigma $, and $\kappa $ (solid lines in Fig. 4a-d) are in good agreement
with experiment\cite{Efimova71} (filled squares) for $T\lesssim 700K$. (The
deviations for $T\gtrsim 700K$ originate in our neglect of the hole
contribution to the charge and heat transport.) In the presence of
nanoinclusions, the individual components of $ZT$ deviate from their bulk
PbTe values at all temperatures shown. For $T\gtrsim 400K$ the increase of
the Seebeck coefficient is compensated by decrease of the conductivity, and
the optimized power factor is close to that of bulk PbTe (see inset in Fig.
4c). At such temperatures the small increase of the $ZT$ factor due to
`electron-only' scattering by nanoinclusions (open circles in Fig 4a) is a
result of the decrease of $\kappa _{e}$ (open circles in Fig. 4d). Comparing
the $ZT$ shown by filled and open circles in Fig. 4a one can conclude that
at a doping $n=2.5\times 10^{19}cm^{-3}$ the enhancement of the optimized $%
ZT $ is primary due to decrease of $\kappa _{ph}$, at least for $T\gtrsim
400K$.

To get a more comprehensive understanding of the role of nanoinclusions in
enhancing the thermoelectric properties, we show in Fig. 5 the
room-temperature $ZT$ factor as a function of the interface potential for
two values of the doping. In addition, we plot $ZT$ calculated using the
bulk value of the phonon thermal conductivity $\kappa _{ph}^{bulk}$ (dotted
lines). We first consider the situation of high doping, as depicted in panel
(a). In the absence of a spatially-varying potential $\left( V_{B}=0\right) $
and without phonon scattering on nanoinclusions, $ZT$ is given by the filled
circle. Turning on the phonon scattering gives a modest $25\%$ increase in $%
ZT$ (the star in the figure). Similarly, one can consider the increase in $%
ZT $ without phonon scattering on the nanoinclusions (dotted line); in this
case, a large increase in $ZT$ of up to $224\%$ is obtained. Thus at this
doping, electron scattering can give a much larger increase in $ZT$.
However, the true advantage of nanoinclusions is realized when \textit{both }%
electron and phonon scattering are included, and the $ZT$ factor can be
increased by as much as $430\%$. This increase is much larger than simply
the sum of the individual electronic and phonon contributions.

The origin of this behavior lies in the non-additive effects of electronic
and phonon thermal conductivities, since $ZT$ depends inversely on their
sum. For the large doping situation of Fig. 5a we have $\kappa
_{e}^{bulk}=4.2W/mK$, $\kappa _{ph}^{bulk}=2.0W/mK,$ and therefore $\kappa
_{ph}^{bulk}<\kappa _{e}^{bulk}$. In this case, reducing $\kappa _{ph}$ by
itself does not lead to an appreciable gain in $ZT$. However, when $\kappa
_{e}$ is also reduced because of scattering and becomes comparable to $%
\kappa _{ph}$, then both work in concert and lead to a large increase in $ZT$%
. Thus, one can imagine that electron scattering on the electrostatic
potential serves as an amplification mechanism to enhance the impact of the
reduction in phonon thermal conductivity. This mechanism works here because
at high doping (1) $\kappa _{ph}^{bulk}<\kappa _{e}^{bulk}$ and (2) the
power factor is maintained or even enhanced in a wide range of interface
potentials.

The situation is quite different in the case of low doping, where $\kappa
_{ph}^{bulk}\gg \kappa _{e}^{bulk}$, as illustrated in Fig. 5b. In this
case, the electronic thermal conductivity is already quite low, $\kappa
_{e}^{bulk}=0.6W/mK<\kappa _{ph}^{bulk}=2.0W/mK$, and the main impact of
nanoinclusions is to decrease the phonon thermal conductivity. The maximum
increase in $ZT$ is $107\%$, with $94\%$ coming from phonons alone. In fact,
for this low doping, the power factor is always reduced compared to the
inclusion-free system -- a signature of this effect is the reduction of $ZT$
below that of the inclusion-free system for larger values of $V_{B}$.

\begin{figure}[h]
\centering
\epsfig{file=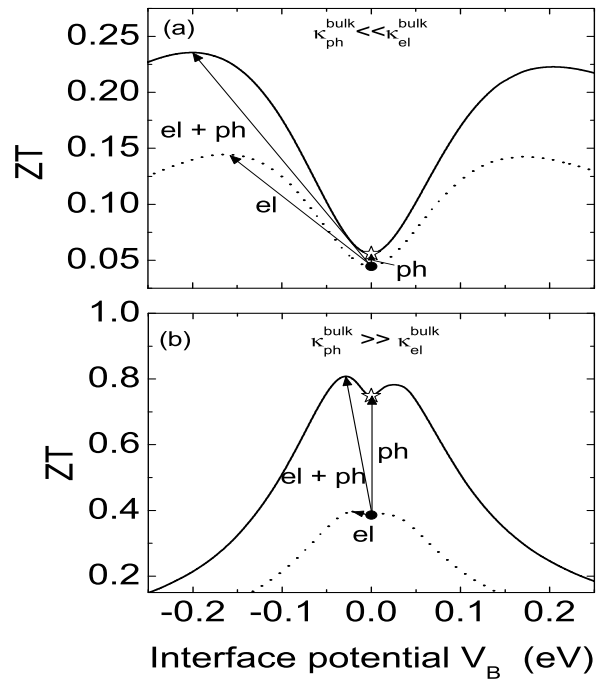,clip=,width=240pt,height=260pt}
\caption{Temperature dependence of the
optimized $ZT$ factor for PbTe. (a) low doping $n=5\times 10^{18}cm^{-3}$
and (b) high doping $n=5\times 10^{19}cm^{-3}$. In both panels filled
circles denote the optimized $ZT$ calculated with both electron and phonon
scattering on nanoinclusions, open circles denote $ZT$ calculated with
electron scattering on nanoinclusions and with bulk PbTe values of $\protect%
\kappa _{ph}$,\ and the solid line is for bulk PbTe. The inset in (a) shows
the values of $V_{B}$ that maximize $ZT$. In (b) the bottom inset shows the
optimal values of $V_{B}$ that maximize $ZT$ (filled circles) and $V_{B}^{Pb}
$ for Pb nanoinclusions. The top inset shows the calculated power factor.
}
\end{figure}

Fig. 6 shows the calculated $ZT$ as a function of temperature for low ($%
n=5\times 10^{18}cm^{-3}$) and high ($n=5\times 10^{19}cm^{-3}$) doping
levels. Included in the figure are the $ZT$ factor calculated with both
electron and phonon scattering on nanoinclusions (filled circles), that
calculated with `electron-only' scattering on nanoinclusions and with bulk
PbTe values of $\kappa _{ph}$ (open circles),\ and the $ZT$ calculated for
inclusion-free bulk PbTe (solid line). The corresponding values of $V_{B}$
that maximize $ZT$ are shown in the insets by filled circles. The inset in
Fig. 6a shows that at $n=5\times 10^{18}cm^{-3}$ the optimal $V_{B}$ is very
small $|V_{B}|<0.03eV$ and even vanishes for $T>600K$. Thus, the electron
contribution to enhancement of optimized $ZT$ is negligible (solid line and
line with open circles almost coincide in Fig. 6a) and the enhancement of
the optimized $ZT$ is dominated by the reduction in $\kappa _{ph}$\ due to
phonon scattering on the inclusions. This can be explained by the fact that
for bulk PbTe the Seebeck coefficient increases with decrease of the doping
concentration $n$, and the \textit{relative} enhancement of the Seebeck
coefficient from its bulk value due to electron scattering on inclusion is
smaller at low doping compared to high doping (see Fig 2). As a consequence
the reduction of $\sigma $ at low doping overweights the increase of $S^{2}$
and the power factor is reduced compared to the inclusion-free system,
leading to small or vanishing optimal $V_{B}$. From a practical point of
view this result means that in order to enhance the $ZT$ factor at low
doping levels one needs to find a metal that gives little or no interfacial
potential.

\begin{figure}[h]
\centering
\epsfig{file=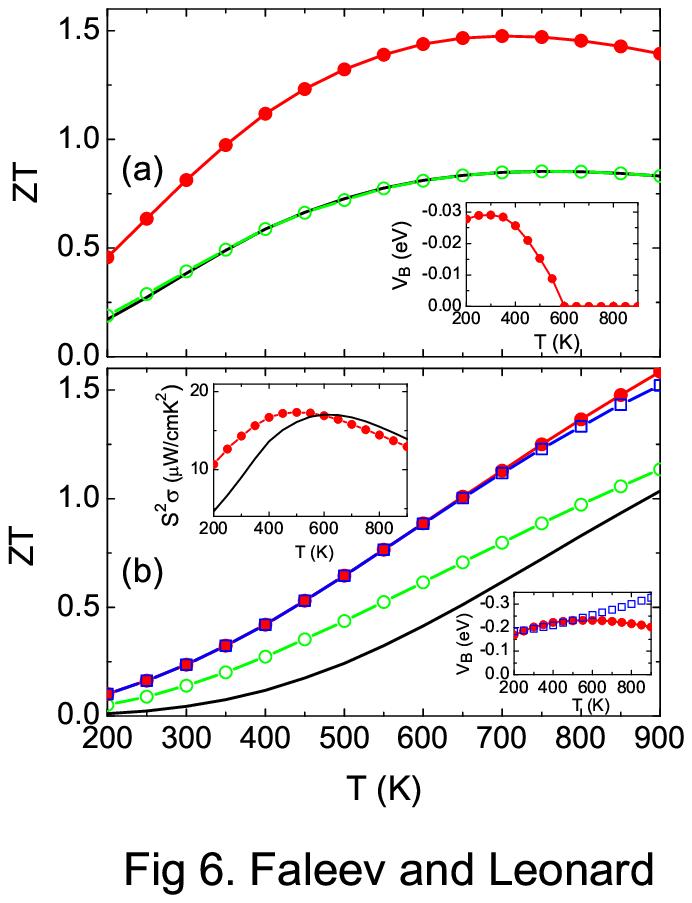,clip=,width=240pt,height=300pt}
\caption{Panel (a) shows the calculated Seebeck
coefficient and conductivity for PbTe as a function of the interface
potential $V_{B}$. Panel (b) shows the resulting power factor. Parameters
are $R=1.5nm$, $T=300K$, $x=5\%$, and $n=2.5\times 10^{19}cm^{-3}$.}
\end{figure} 

For larger doping the electron contribution to enhancement of $ZT$ becomes
important and the optimized $V_{B}$ increases. It is seen in Fig. 6b \ that
for $n=5\times 10^{19}cm^{-3}$ the `electron-only' contribution to
enhancement of the optimized $ZT$ makes up over 50\% of the enhancement at $%
T\lesssim 600K$ and the optimized $V_{B}$ is as large as $0.2eV$. Moreover,
at large doping, $\kappa _{e}^{bulk}>\kappa _{ph}^{bulk}$, and the reduction
of $\kappa _{e}$ due to scattering on inclusions amplifies the effect of the
reduced $\kappa _{ph}$. The upper inset in Fig. 6b shows that the power
factor $\sigma S^{2}$ is enhanced only for $T<600K$. The reduction of the
power factor relative to the inclusion-free system at $T>600K$ is due to the
fact that at high temperature the Seebeck coefficient of bulk PbTe is
already large (see Fig. 4b), the \textit{relative} increase in $S$ induced
by electron scattering on inclusions becomes smaller at increased
temperature, so the reduction of $\sigma $ overweights the increase of $S^{2}
$. Nevertheless, the `electron-only' contribution results in enhancement of $%
ZT$ (open circles in Fig. 6b) even at higher temperatures due to reduction
of $\kappa _{e}$.

Figure 6b shows by open squares the $ZT$ factor for PbTe with Pb inclusions
(assuming $V_{B}^{Pb}-E_{F}=-0.35eV$) for parameters $n=5\times
10^{19}cm^{-3}$, $R=1.5nm$, and $x=5\%$. For this set of parameters the
interface potential $V_{B}^{Pb}$ is close to the optimal one (see inset in
Fig 6b) in a wide range of temperatures, so the $ZT$ factor for the system
with Pb inclusions is very close to the optimal $ZT$. The enhancement of $ZT$
due to Pb inclusions is on the order of 400\% at room temperature and 50\%
at $T=900K$, where it reaches a value as high as $1.5$.

\section{Conclusion}

In conclusion, we developed a theory that allows the calculation of the $ZT$
factor and its components for a system of a semiconductor host material with
spherical metallic nanoinclusions. The enhancement of the Seebeck
coefficient can be explained by a strong energy dependence of electron
scattering on the band-bending at the interface between metallic inclusions
and the semiconductor host. The electronic contribution to enhancement of $%
ZT $ is important for high doping, while at low doping the enhancement of $%
ZT $ is dominated by the reduction in the phonon thermal conductivity.\ The
theory can be used to choose the optimal parameters for the metal
nanoinclusions (interface potential, inclusion volume fraction or radius) in
order to maximize $ZT$.

We thank Doug Medlin and Peter Sharma for useful discussions. Sandia is a
multiprogram laboratory operated by Sandia Corporation, a Lockheed Martin
Company, for the United States Department of Energy under contract
DE-AC04-94-AL85000.

\end{document}